\documentstyle[11pt,aaspp4,epsf,rotate]{article}
\eqsecnum
\def\beq{\begin{equation}}
\def\eeq{\end{equation}}
\def\simge{\mathrel{%
   \rlap{\raise 0.511ex \hbox{$>$}}{\lower 0.511ex \hbox{$\sim$}}}}
\def\simle{\mathrel{
   \rlap{\raise 0.511ex \hbox{$<$}}{\lower 0.511ex \hbox{$\sim$}}}}
\begin{document}
\title{Cross-Correlation of the $2-10~keV$ XRB with Radio Sources: 
Constraining the Large-Scale Structure of the X-ray Background}
\author{Stephen P. Boughn$^{1,2}$}
\affil{$^1$Institute for Advanced Study, Olden Lane, Princeton, 
NJ 08540}
\affil{$^2$Department of Astronomy, Haverford College, Haverford, PA  19041
sboughn@haverford.edu}

\begin{abstract}
We present cross-correlation analyses of the HEAO $2-10~keV$ diffuse X-ray map
with both the combined GB6/Parkes-MIT-NRAO (GB6-PMN) $5 GHz$ and the 
FIRST $1.4 GHz$ radio surveys.
The cross-correlation functions (CCFs) of both radio surveys with the unresolved
X-ray background were detected at the $5 \sigma$ level.  While the large angular 
resolution ($3^\circ$) of the X-ray map makes it difficult to separate the 
contributions of clustering from those of Poisson fluctuations, the amplitude of
the CCF provides important constraints on the X-ray emissivity of the radio 
sources as well as on the clustering properties of radio and X-ray sources.  These
constraints are subject to a number of modeling parameters, e.g. X-ray 
luminosity evolution, clustering evolution, the radio luminosity function,
cosmological model, etc.  For reasonable choices of paramters the
X-ray/FIRST CCF is consistent with a correlation scale length of $6h^{-1} Mpc$.  This is
somewhat smaller than the scale length inferred from the autocorrelation function
of the FIRST survey and implies that X-ray sources are less strongly clustered 
than strong radio sources, a result which is consistent with previous constraints 
on X-ray clustering.  The X-ray/GB6-PMN CCF is several times larger and is likely 
to be dominated by Poisson fluctuations.  This implies that $\sim 2\%$
of the diffuse X-ray background arises from the GB6-PMN sources.

\end{abstract}

\keywords{diffuse radiation $-$ galaxies:statistics $-$ large-scale structure of 
the universe $-$ X-rays: galaxies $-$ X-rays: general}

\section{Introduction}

Thirty-five years after the discovery of the cosmic X-ray background (XRB), it is 
still the subject of a great deal of study.  On the one hand it offers the 
possibility of providing an extremely useful tool for the study of large scale 
structure in the universe (\cite{bfc97}; \cite{bct97})
and yet on the other the nature and origin of the XRB is still 
not well understood.  The deep images made by the ROSAT satellite have resolved 
$60\% $ of the $0.5-2~keV$ background into discrete sources (\cite{has93})
and similar observations by the ASCA satellite have resolved 30\% of the $2-10~keV$
background (\cite{geo97}).  While it is clear that classical 
active galactic nuclei (AGN), i.e. QSO's, make a significant contribution to 
the XRB (\cite{geo97}; \cite{boy94}), it is also clear for a variety of reasons that
a substantial contribution must come from some other population.  
$2-10~keV$ number counts are a factor of $2$ to $3$ larger than inferred from 
$0.5-2.0~keV$ counts if one assumes a typical AGN X-ray spectrum 
(\cite{geo97}).  Indeed, the spectrum of the XRB is significantly harder 
that that of AGN (\cite{gen95}).  Finally, the strong clustering of
QSO's is inconsistent with the relatively smooth XRB (\cite{geo97} and references
cited therein).
These observations point to a large population of relatively faint (or highly
absorbed) sources with hard X-ray spectra.  Such sources have already begun to
be identified with faint galaxies and X-ray bright, narrow emission line galaxies
(\cite{alm97}; \cite{af97}; \cite{geo97}; \cite{rhm97}; \cite{tl96}; \cite{roc95}).
On the other hand, Comastri et al. (1995) have successfully reproduced the flux
and spectrum of the XRB with a model AGN luminosity function that includes a 
large number of highly absorbed Seyfert 2s.  In either case, if such sources make
a substantial contribution then one might expect the X-ray background to be 
clustered more like galaxies than QSOs.  This is consistent with the results of the
cross-correlation analysis presented below.

In this paper we undertake cross-correlation analyses of the `hard' ($2-10~keV$)
X-ray background with two flux limited radio source surveys, the FIRST survey at
1.4 GHz and the combined GB6/Parkes-MIT-NRAO (GB6-PMN) surveys at 4.85 GHz.  While the
flux limits of these surveys differ by a factor of $\sim 40$, their expected redshift
distributions are similar and, therefore, the cross-correlation functions of the
two surveys can be directly compared.  The description of the data sets (X-ray 
and radio) and the editing of these sets is described in \S2.  The 
cross-correlation analysis and significance tests are decribed in \S3.  
The formalism to interpret the CCF in terms of a model follows closely the 
analysis of Treyer and Lahav (1996) and is presented in \S4.  The
constraints on parameters resulting from the observed CCF's are discussed in \S5.

\section{Data Sets}

\subsection{HEAO1 A2 ${\mathbf 2 - 10~keV}$ X-ray Map}

The HEAO1 A2 experiment measured the surface brightness of the X-ray
background in the $0.1 - 60~keV$ band (\cite{bol87}).  The present data set was
constructed from the output of two medium energy detectors (MED) with different
fields of view ($3^\circ \times 3^\circ$ and $3^\circ \times 1.5^\circ$) and 
two high energy detectors (HED3) with
the same fields of view.  These data were collected during the six month period
beginning on day 322 of 1978.  Counts from the four detectors were combined and 
binned in 24,576  $1.3^\circ \times 1.3^\circ$ pixels in an equatorial quadrilateralized 
spherical cube projection on the sky (\cite{ws92}).  The combined map has a
spectral resolution of approximately $2 - 10~keV$ (\cite{jm89}).
The effective point spread function (PSF) of the map was determined by averaging 
the PSF's of 75 HEAO1 point sources (\cite{pin82}).  The composite PSF 
is well fitted by a gaussian with a full width, half maximum of $2.96^\circ$.  
Because of the pixelization, the PSF varies
somewhat with location on the sky; however, this has little effect on the 
correlation analysis and so a constant PSF is used in the analysis of \S4 below.

The dominant feature in the HEAO map is the Galaxy, so all data within $20^\circ$ of
the Galactic plane and within $30^\circ$ of the Galactic center were cut from the map.
In addition, $10^\circ$ diameter regions around $90$ discrete X-ray sources with
$2 - 10~keV$ fluxes larger than $3 \times 10^{-11} erg s^{-1} cm^{-2}$ (\cite{pin82})
were removed.  Without this cut the CCF's were somewhat larger and considerably
more noisy due to bright, nearby Galactic and extragalactic sources.  The
resulting ``cleaned'' map covered about 50\% the sky.  In order to identify
additional point sources, 
the map itself was searched for ``sources'' that exceeded the nearby background
by a specified amount and $7^\circ$ diameter regions around these were removed.  Cuts 
were made at several levels from 4 to 10 times the photon noise.  For
the most extreme cuts which corresponded to a point source flux of 
$3 \times 10^{-11} erg s^{-1} cm^{-2}$ the sky coverage was reduced to about 25\% of 
full sky.  The results
of these additional cuts did not significantly affect the correlation analyses 
and we conclude that the X-ray map contains no point sources with fluxes greater
than $3 \times 10^{-11} erg s^{-1} cm^{-2}$.  This flux cut is accounted for in 
the analysis of \S4.

Even after cleaning, the X-ray map has several components of large-scale
systemic structure which can be corrected for.  If the dipole moment of the
cosmic microwave background is a kinematic effect, as it has been widely
interpreted (\cite{ben96}), then the X-ray background should possess
a similar dipole structure (Compton-Getting effect) with an amplitude of
$4.3 \times 10^{-3}$.  Evidence for this structure is, indeed, found in the HEAO map
(\cite{sha83}; \cite{lpt97}).  The cleaned map was corrected for this effect.
In addition, a linear time drift in detector sensivity (\cite{jah93}) results
in a large scale structure of known form.   Finally, the $2 - 10~keV$
background shows evidence of high latitude Galactic emission as well as emission
associated with the Supergalactic plane (\cite{jah93}).  Models for these 
contributions along with the time drift were fit to the X-ray data and 
subseqently subtracted from the map (\cite{bct97}).
These contributions to the X-ray background are on large scales and
have little effect on the small angular scale correlation analysis 
discussed below.

Because of the ecliptic longitude scan pattern of the HEAO satellite, sky
coverage and, therefore, photon shot noise are not uniform.  However, the
mean variance of the cleaned, corrected map, $2.0 \times 10^{-2} (cts/sec)^2$, is
considerably larger than the mean variance of photon shot noise, $0.67 \times 10^{-2}
(cts/sec)^2$, where $1 cts/sec = 2.1 \times 10^{-11} erg s^{-1} cm^{-2}$ (\cite{ajw94}).
This implies that the X-ray map is dominated by ``real'' 
structure (not photon shot noise).  For this reason, in the correlation
analyses that follow, we chose to weight each pixel equally.

\subsection{FIRST 1.4 GHz Survey}

The FIRST $1.4 GHz$ survey is a continuing project to survey $10,000$ square 
degrees of the north Galactic cap (\cite{wbhg97}).
The data used in the analysis below was obtained from the publically available
catalog containing $236,177$ sources from observations of 1993 through 1996.
The catalog covers about $2575$ square degrees and includes only sources whose
peak flux exceeds $5$ times the rms noise plus $0.25\  mJy$.  Following Cress et al.
(1996), all pairs of sources within $0.02$ degrees of each other are considered
to be part of a single doubled lobed source.  For groups of three or more such
sources, all sources within $0.02$ degrees of the mean position of the group 
are considered to be part of a multi-component system and are
counted as a single source.  This reduces the total 
number of sources to $186,214$.  Since noise is not uniform across the 
coverage region, especially in those areas near very bright sources, the 
flux limit is not uniform.  To correct for this we have flagged all map areas in
which the rms noise is greater than $0.17  mJy$ and have removed all sources with
peak fluxes less than $1.1  mJy$ which corresponds to a $5 \sigma$ detection if 
$\sigma = 0.17  mJy$.  This further reduces the number of sources to $163,157$.  It
should be noted that even without the latter correction, the CCF of FIRST 
sources with the X-ray background is not changed significantly.  This is 
understandable since non-uniform coverage in the FIRST catalog is not expected
to be correlated with systematic structure in the X-ray map.

The remaining FIRST sources are grouped in the same $1.3^\circ \times 1.3^\circ$
pixels of the
quadrilateralized cube projection used for the X-ray map.  Because of flagged
regions as well as the projection itself, not all pixels represent the same
solid angle coverage of radio sources.  Therefore, the radio coverage of each
pixel is used to weight its contribution to the cross-correlation in a way so as not
to bias the result.  See \S3 below.  The number of pixels that contain
FIRST data is $1696$; although, somewhat less, $1100$, are common to both the FIRST
and X-ray maps.

\subsection{Parkes-MIT-NRAO and GB6 4.85 GHz Surveys}

The publically availible Parkes-MIT-NRAO (PMN) southern sky survey was made 
with the 64-m radio telescope 
at Parkes, NSW, Australia, and contains about $50,000$ sources (\cite{wri94}; 
\cite{gri94}; \cite{gri95}; \cite{wri96}).  The flux limit in this combined 
survey is not uniform but varies from
$20\ mJy$ to $72\ mJy$.  As a compromise between uniform coverage and total number of 
sources we chose to use a $50\ mJy$ flux limited sample and flagged all portions of the sky
not covered to that level.  This required excluding the Zenith zone of the survey.
In addition, only those sources with $Dec < 0^\circ$ were included since the northern 
sources overlapped with the GB6 survey.  Following Loan, Wall, \& Lahav (1997) we
excluded several small regions with extended sources.  Finally all regions of the sky
within $20^\circ$ of the Galactic plane and within $30^\circ$ of the Galactic center
were removed from consideration.  After these cuts 15,233 sources remain in the catalog.

The Green Bank 6cm (GB6) survey of the northern sky ($0^\circ < Dec < 75^\circ$)
was made with the NRAO 91m telescope during 1986-7 and contains $75,162$ sources brighter
than $\sim 18 mJy$ (\cite{gsdc96}).  Rather than worry about comparing the flux
calibrations of the GB6 and PMN surveys we chose a somewhat smaller
flux limit for the GB6 survey,
$45  mJy$, which resulted in the equality of the surface density of sources
($1.50  deg^{-2}$) in the two maps.  In any case, the correlation analysis below was
performed on the two surveys separatedly as well as on the combined GB6-PMN survey.

The combined GB6-PMN flux limited sources were also grouped in the same $1.3^\circ 
\times 1.3^\circ$ pixels as the X-ray map.  There are $\sim 34,000$ sources and 
$13,520$ pixels in the combined map which corresponds to $55\% $ sky coverage;
 although, only $10,115$ pixels are 
common to both the GB6-PMN and cleaned X-ray maps.

\section{Cross-Correlation Functions}

We define the dimensionless cross-correlation function of the X-ray intensity, $I$, 
with the radio source number, $N$, as
\begin{equation}
W(\theta)_{I,N} = \sum_{i,j} (I_i -\bar{I})(N_j - \bar{N_j})/
\sum_{i,j} \bar{I}\bar{N_j}
\end{equation}
where the sum is over all pairs of pixels, $i,j$, separated by an angle $\theta$, $\bar{I}$
is the mean X-ray intensity, and $\bar{N_j}$ is the mean number of radio sources in the $j^{th}$
pixel.  As discussed in \S2.2, the FIRST radio coverage of each $1.3^\circ \times 1.3^\circ$
pixel is not the same.  Therefore, we take $\bar{N_j}=\bar{n}\Omega_j$ where $\bar{n}$ 
is the mean surface density of radio sources and $\Omega_j$ is the solid angle of radio
coverage of the $j^{th}$ pixel.

Figures 1 and 2 are the CCFs of the $2-10~keV$ HEAO map with the combined GB6 and
PMN surveys and with the FIRST survey.  Although, the CCFs are shown out to separation
angles of $15^\circ$, a signficant signal is only detected in the first few bins.  The errors
were computed using a ``bootstrap'' analysis (\cite{cre96}; \cite{fis94}; \cite{lfb86}).  
100 random radio 
source catalogs, each of the same size as the original catalog were generated by choosing
sources at random from the original catalog.  Note that this requires some sources
will be chosen more than once.  These random catalogs are then cross-correlated with the
real X-ray map according to equation (3-1).  The mean CCF of the random trials was consistent
with that of the real data and the rms fluctuation about the mean CCF provides an estimate
of the uncertainty due to the additional Poisson noise in the distribution of radio sources.

For the FIRST/X-ray CCF, the error estimates where checked in two ways.  
A series of 59 radio source 
maps were generated by a reflection through the celestial equator followed by a rotation
about the celestial pole in $6^\circ$ incremants.  These radio maps were then 
cross-correlated with the X-ray map which was first transformed into Galactic coordinates.
This latter transformation resulted in a pixel coverage  nearly the same as that
of the original data for small separation angles.  The resulting transformed 
X-ray and radio maps possessed little small-scale correlation.  The rms scatter of
the CCF's of this set of maps agreed with the bootstrap error estimates for 
$\theta < 3^\circ$ and was about $50\%$ larger than the bootstrap estimates for
$\theta > 3^\circ$.  An additional rough error estimate was obtained by dividing 
the two data sets in half and comparing the CCFs of both halves.  For two 
different partitions, north-south and
east-west, the differences in the two CCFs were consistent with the quoted errors.
Both these estimates imply that the bootstrap error estimates are reasonable. 
We consider the bootstrap error estimates preferrable in that 
they reflect the actual distribution of data on the sky. 
To compare the consistency of the GB6 and PMN portions of the $5GHz$ map, the CCF
was computed separately for each and the results are plotted in Figure 3.
It is clear that they are consistent with each other (and with Figure 1) to within 
the estimated errors.

Figures 1 and 2 demonstrate that there is a statistically significant cross-correlation 
of radio source counts and the $2-10~keV$ background at the $5\sigma$ level.  It
may appear from the figures that the significance is higher than this; however, 
because of the X-ray PSF, the error bars are highly correlated.  The correlation
coefficients of adjacent errors are typically between $0.4$ and $0.7$.

Also evident in Figures 1 and 2 is that for $\theta \leq 2^\circ$ the GB6-PMN/X-ray CCF
exceeds the FIRST/X-ray CCF by a factor of $\sim 3$.  In addition, it appears that the 
latter CCF is more extended than the former.  It is possible that these two
properties are related.  It will be shown in \S4 that the finite PSF 
of the X-ray map results in a 
$W(\theta)$ profile similar to that of the GB6-PMN/X-ray CCF even if the only
cross-correlation arises from the Poisson noise in individual sources.  To the extent
that the GB6-PMN/X-ray CCF is dominated by Poisson noise it will be both larger in 
amplitude and more narrow in angular scale than a CCF for which Poisson noise
is negligible.  We suggest in \S4 that this is the case here.

\section{Interpretation of the Cross-Correlation Function}

The observed $W(\theta )$'s in \S3, depend on the properties of radio sources and
X-ray sources, the spatial clustering of these populations, and the large scale 
geometry of the universe.  Among the quantities included in the following model 
of $W(\theta )$ are the luminosity function of the radio sources, the luminosity (and
density) evolution of radio sources, the X-ray luminosity of radio sources,
the spectrum  and evolution of X-ray emissivity,
the functional form and evolution of the spatial cross-correlation function, and the
cosmological parameters, $H_\circ$ and $q_\circ$.  Although these parameters provide
considerable freedom in fitting the observed $W(\theta )$, the constraints placed on
parameter space are reasonably strong.  The analysis of this section follows 
closely that of Treyer and Lahav (1996).  The reader is referred to that paper for
a detailed analysis.

Let $\eta_\circ$ be the unnormalized, angular cross-correlation function.  If the
sky is divided up into cells of small solid angle $\omega$, then
\begin{equation}
\eta_\circ  = \langle \delta N\delta I\rangle=
\langle (N-\langle N\rangle)(I-\langle I\rangle)\rangle=
\langle NI\rangle -\langle N\rangle \langle I\rangle
\end{equation}
where $N$ and $I$ are the number of radio sources in and average X-ray intensity
of each cell, and the average is over all cells.  
It is straightforward to show that (\cite{tl96}; \cite{peb80})  
\begin{equation}
\langle NI\rangle=\int {\varepsilon_r \over 4\pi r_{L}^{2}} dV +
\int \int n({\bf r}_1){\varepsilon_b ({\bf r}_2)\over 4\pi r_{L_2}^{2}}
[1+\xi(r_{12})]dV_1dV_2
\end{equation}
and
\begin{equation}
\langle N\rangle \langle I\rangle = \int n({\bf r}_1)dV_1\int 
{\varepsilon_b ({\bf r}_2)\over 4\pi r_{L_2}^{2}} dV_2
\end{equation}
where $\varepsilon_r$ and $\varepsilon_b$ are the comoving 
volume X-ray emissivities of the
radio source population and the total X-ray background respectively,
$n$ is the comoving number density of radio sources, $r_L$ is the luminosity 
distance, and the integrals are performed over the comoving volumes subtended 
by the solid angle $\omega$ of the cell.  $\xi (r_{12})$ is the spatial 
cross-correlation function of radio sources with the X-ray background 
in regions of space separated by a proper distance $r_{12}$, i.e.
\begin{equation}
\langle n({\bf r}_1)\varepsilon_b({\bf r}_2)\rangle = 
[1+\xi(r_{12})]\langle n\rangle \langle \varepsilon_b \rangle
\end{equation}
Thus $\eta_\circ = \eta_P + \eta_c$ where
\begin{equation}
\eta_P =\int {\varepsilon_r \over 4\pi r_{L}^{2}} dV
\end{equation}
and
\begin{equation}
\eta_c = \int \int n({\bf r}_1){\varepsilon_b ({\bf r}_2)\over 4\pi r_{L_2}^{2}}
\xi(r_{12})dV_1dV_2 
\end{equation}
The first term, $\eta_P$, arises from Poisson fluctuations
due to the X-ray emission of the individual radio sources and is equal to the X-ray
flux from these sources, i.e.,
\begin{equation}
\eta_P = \omega \bar{I_r}
\end{equation}
where $\bar{I_r}$ is the mean X-ray intensity of the radio sources.
The second term, $\eta_c$, is due to the joint clustering of radio sources
with the sources of the X-ray background (including the radio sources).

The spatial auto-correlation function (ACF) of nearby galaxies is well approximated by
a power law.  We take this form for the spatial cross-correlation function 
and assume the standard power law evolution (\cite{peb80})
\begin{equation}
\xi (r,z) = (1+z)^{-(3+\epsilon)}\left({r\over r_\circ}\right)^{-\gamma}
\end{equation}
where $r$ is the proper (non-comoving) distance between the sources, $r_\circ$ is the comoving
correlation length, and $\epsilon$ is a clustering evolution parameter.  For
`stable' clustering $\epsilon = 0$ while $\epsilon = \gamma - 1$ for linearly
growing perturbations in an Einstein-de Sitter universe (\cite{tl96}).
Because the correlation scale length $r_\circ$ is small, i.e. $\sim 10Mpc$, sources
at significantly different redshifts are uncorrelated.  In this case the integrals 
in equation (4-6) can be simplified to a single integral over redshift (\cite{tl96}),
\begin{equation}
\eta_c = K_\gamma H_\gamma r_{\circ}^{\gamma} \int {n(z)\varepsilon_b(z)\over 4 \pi
r_{L}(z)^{2}} (1+z)^{-(3+\epsilon )+\gamma} r_c(z)^{5-\gamma}F(z)dr_c(z)
\end{equation}
where $K_\gamma = \int d\Omega_1 \int d\Omega_2 {\theta_{12}}^{1-\gamma}$;
 $\theta_{12}$ is the angle between ${\mathbf\theta}_1$ and ${\mathbf\theta}_2$; 
$H_\gamma = \Gamma ({1\over 2})\Gamma ({\gamma -1\over 2})/
\Gamma({\gamma \over 2})$; $r_c$ is the comoving radial coordinate; and 
$F(z){r_c}^2 dr_c d\Omega \equiv dV$.  This expression can be further simplified by
noting that $r_L (z) = (1+z)r_c (z)$ and $n(z)dV/d\Omega = n(z)F(z){r_c}^2 dr_c
=N(z)dz$ where $N(z)dz$ is the surface density of radio galaxies at redshift $z$.
Following Treyer \& Lahav (1996) we assume a power law evolution of the 
${\it observed}$ XRB volume emissivity, i.e. 
$\varepsilon_b (z) = \varepsilon_{b,\circ}(1+z)^q$.  Note that $q$ includes the 
``K-correction'' exponent, $1+\alpha$, where $\alpha$ is the energy spectral 
index.  Then the expression for $\eta_c$ becomes
\begin{equation}
\eta_c = {K_\gamma H_\gamma {r_\circ}^\gamma \varepsilon_{b,\circ}\over 4\pi}
\int N(z)(1+z)^{\gamma +q-5-\epsilon}r_c (z)^{1-\gamma}dz
\end{equation}
Equations (4-7) and (4-10) give the value of the cross-correlation at zero
separation for an X-ray map with a delta function PSF.  It is straightforward 
to show that for a finite PSF and for arbitrary separation angle, $\theta$,
$\eta_P$ becomes
\begin{equation}
\eta_P (\theta) = B(\theta)\bar{I_r}
\end{equation}
where 
\begin{equation}
B(\theta) = \int_{radio} P({\mathbf\theta}-{\mathbf\theta^\prime})
d\Omega^\prime ,
\end{equation}
$P(\theta^\prime)$ is the normalized PSF,
i.e., $\int P(\theta^\prime)d\Omega^\prime = 1$, the integration is over the radio cell,
and ${\mathbf\theta}$ is the location of the X-ray cell relative to the radio cell.  
The expression for 
$\eta_c (\theta)$ is again equation (4-10) if $K_\gamma$ is substituted with
\begin{equation}
K_\gamma (\theta) = \int_{S} d\Omega_1 P(\theta_1)\int_{radio} d\Omega_2
{\theta_{12}}^{1-\gamma}
\end{equation}
where $\int_{S}$ indicates an integral over all space.  These two expressions
are equivalent to those derived by Refregier, Helfand, and McMahon (1997) recalling
that the effective PSF used here is the actual PSF averaged over an X-ray cell.
It should be noted that, due to the finite PSF, the Poisson term contributes 
to the CCF for $\theta > 0$.  Of course, this is why it is problematic to
distinguish real clustering from Poisson fluctuations.  

As an illustration, 
Figures 1 and 2 show a fit of $B(\theta)$ to the GB6-PMN/X-ray CCF
and a fit of $K_\gamma (\theta)$ to the FIRST/X-ray CCF.  
We have chosen $\gamma=2.0$ to evaluate $K_\gamma$ which
is consistent with the auto-correlation function of FIRST sources found by
Cress et al. (1996) and similar to the value ($\gamma = 1.8$) for local bright
galaxies (\cite{peb80}).  Weighted least squares fits to the first 4 data points
were performed following a similarity transformation to diagonalize the noise
matrix.  As long as the first three points are included, the results are rather 
insensitive to the number of points included.  Because of the pixelization of the
data, theoretical functions $K_\gamma (\theta)$ and $B(\theta)$ are evaluted at 
only those angles appropriate for the data.  For aesthetic reasons, these points 
are connected by straight lines in the Figures.
Both of the curves look reasonable for $\theta < 5^\circ$.  For $\theta > 5^\circ$,
the data of Figure 2 fall consistently below the theoretical curve.
This is not unexpected since $5^\circ$ corresponds to rather large distances at even
modest redshifts, e.g. at $z=0.2$ an angle of $5^\circ$ corresponds to $44h^{-1}Mpc$
for an Einstein-de Sitter universe.  On the other hand the observed galaxy ACF
displays a break below the power law at lengths $\simge 30h^{-1}Mpc$ (\cite{peb93}).
This behavior is consistent with the evolution of very large scale structure in
a standard CDM universe (see e.g. \cite{pad93}). 
As an indication of the magnitude of this
effect we have constructed a theoretical CCF from equation (4.6) with the power-law
$\xi(r_{12})$ cutoff above $r_{12} = 30h^{-1}Mpc$.  Figure 4 is a fit of this profile to
the FIRST/X-ray CCF.
The discrepancy at large angles is no longer egredious while the amplitude of the fit
is nearly the same as in Figure 2.  We make no claim that this model has any 
particular significance but offer it as an indication of the magnitude of the effect.

The formal ${\chi_\nu}^2$ for the two fits in Figures 1 \&  2 
are $0.2$ and $1.7$ respectively for three degrees of freedom.  
On the other hand, fits of $K_\gamma (\theta)$ to the GB6-PMN/X-ray
CCF and of $B(\theta)$ to the FIRST/X-ray CCF  give ${\chi_\nu}^2 = 6.2$ and $7.3$.  
Fitting the 
$B(\theta)$ and $K_\gamma (\theta)$ profiles simultaneously to the two CCFs
doesn't improve ${\chi_\nu}^2$ and, in fact, is consistent with no $B(\theta)$
contribution to the FIRST data and no $K_\gamma (\theta)$ contribution to the
GB6-PMN data.  These results are suggestive
that the GB6-PMN CCF is Poisson dominated while the FIRST CCF is clustering 
dominated.  The conclusions are not overly sensitive to
$\gamma$.  If $\gamma$ is chosen to be that found for the ACF of nearby 
galaxies (\cite{peb80}), i.e., $\gamma = 1.8$, the fit of $K_\gamma$ to the FIRST
CCF is somewhat worse (${\chi_\nu}^2 = 2.3$) while the $K_\gamma$ profile is even
more inconsistent (${\chi_\nu}^2 = 9.4$) with the the GB6-PMN CCF.  
These matters will be discussed further in \S5.

In order to compare the amplitudes of the predicted $\eta_\circ (\theta)$ with 
the observed $W(\theta)$s, a number of parameters must be specified: $\gamma$,
$q$, $\varepsilon_{b,\circ}$, $\epsilon$, $r_\circ$, and $z_{cutoff}$, the
redshift at which X-ray sources ``turn on'' and thus the upper limit to 
the integral in equation (4-10).  The functions $r_c(z)$ 
and $F(z)$ require an assumption about the large-scale geometry of the universe,
and the distribution of radio sources $N(z)$ depends on the evolving
luminosity function of radio sources as well as on the geometry of the universe.

In the analysis that follows $N(z)$ is computed from the fundamental ``free-form
model'' of the radio luminosity function of Dunlop \& Peacock (1990) with 
low flux cutoffs appropriate to the 
two surveys. Figures 5 and 6 
are the $N(z)$s computed from this model for the GB6-PMN and FIRST surveys.
For the $1.1mJy$ peak flux cut in the FIRST survey, the effective completeness flux
is about $1.5mJy$ (\cite{bwh95}; \cite{wbhg97}).
We have repeated the analyses using the evolving luminosity and number density
model of Dunlop \& Peacock and, although the $N(z)$s are somewhat different, 
the differences in the correlation analyses are small.  Because $N(z)$
falls off at large redshift, the computed values of $\eta_c$ are rather
insensitive to $z_{cutoff}$; however, this parameter is important in 
constraining X-ray emissivity.

The emissivity $\varepsilon_{b,\circ}$ of the X-ray background must satisfy 
the constraint that the integrated intensity equal that observed for the
$2-10~keV$ background, i.e. $5.1 \times 10^{-8} erg s^{-1} cm^{-2} sr^{-1}$ 
(\cite{mar80}).
For a given evolution parameter $q$ and cosmological model, 
$\varepsilon_{b,\circ}$ is uniquely determined.  In principle, $q$ could be
determined uniquely from the observed local X-ray emissivity;
however, uncertainty in this value as well as uncertainty in $z_{cutoff}$
result in considerable uncertainty in $q$.  In addition, it is quite likely
that a simple power law evolution is not the best description of X-ray emissivity.
If one arbitrarily sets $z_{cutoff}=4$, then, for an Einstein-de Sitter universe,
the local X-ray emissivities implied for $q=$ $2$, $3$, $4$ are $18.7$, $8.3$, and
$3.0 \times 10^{38}h$ $erg s^{-1} Mpc^{-3}$ where $h=H_\circ/100 km s^{-1}
Mpc^{-1}$.  The locally measured value is
$8.6 \pm 2.4 \times 10^{38}h$ $erg s^{-1} Mpc^{-3}$ for AGN alone with an upper limit
of $4 \times 10^{38}h$ $erg s^{-1} Mpc^{-3}$ for the contribution of weaker sources
(e.g. star-forming galaxies, LINERS) (\cite{miy94}).  It appears that for power law 
evolution models, $q$ is constained to fall between $2$ and $4$.  As a somewhat
more sophisticated model of evolution we consider the unified AGN
model of Comastri et al. (1995) which reproduces both the amplitude and spectrum
of the XRB.  Figure 7 is a plot of the redshift distribution of the X-ray intensity,
${\mathcal{F}}(z) \equiv dI/dz$, from the Comastri et al. model with the flux cut
$3\times 10^{-11}~erg~s^{-1}~cm^{-1}$ appropriate for the present X-ray map (\cite{bct97}).
Expressed in terms of $\mathcal{F}$, equation (4-10) must be modified slightly,
\begin{equation}
\eta_c = {K_\gamma H_\gamma {r_\circ}^\gamma }
\int {{\mathcal{F}} (z)N(z)\over F(z)dr_c/dz}(1+z)^{\gamma -3-\epsilon}
r_c (z)^{1-\gamma}dz .
\end{equation}
For the special case of an Einstein-de Sitter universe ($\Omega_\circ = 1$) 
this equation becomes
\begin{equation}
\eta_c = {K_\gamma (\theta) H_\gamma {r_\circ}^\gamma \over 2^{\gamma -1}
(c/H_\circ )^\gamma}
\int {\mathcal{F}} (z)N(z)(1+z)^{\gamma -{3\over 2}-\epsilon} r_c (z)
[1+(1+z)^{-{1\over 2}}]^{1-\gamma} dz
\end{equation}
where $H_\circ$ is Hubble's constant and $c$ is the speed of light.

The $\eta_c$ computed using this model is intermediate between those
of the $q=3$ and $q=4$ power law models.  We note that the redshift distribution 
of the X-ray $2-10~keV$ intensity for the Comastri et al. model is relatively flat 
and that a $q=3.5$ power law evolution gives a flat redshift 
distribution.  In the analysis that follows, we will use the Comastri et al. model
for the evolving X-ray emissivity as our ``best guess'' but will compare the 
results of this model with those of $q=2,3,4$ power law models.

Because of the large angular resolution ($3^\circ$) of the HEAO map, most of the
contribution to the CCF arises from nearly linearly evolving structures.  At $z=0.2$,
$1^\circ$ corresponds to $9h^{-1}Mpc$ which is comparable to the transition
from the linear to non-linear regime in the local universe.  In the ``best 
guess'' model below, roughly $50\%$ of the contribution to the CCF comes from
redshifts below $z=0.2$ where evolution is modest and $50\%$ comes from
redshifts $z>0.2$ where $1^\circ$ is in the linear regime.  For this
reason we assume for our ``best guess'' model that structure is growing linearly
in an Einstein-de Sitter universe, i.e., $\epsilon=\gamma-1$.  However, the change 
in the results if the clustering is stable, i.e, $\epsilon=0$ and for open and
``$\Lambda$'' universes will be discussed.

Finally one must take into account the flux limit ($3 \times 10^{-11}erg s^{-1}
cm^{-2}$) on X-ray sources.  There is no well-defined procedure to do this for
the power law models of emissivity evolution since individual source
luminosities are not
specified.  However, the flux cut can be directly applied to the Comastri et al.
model and we have done so.  The net result is to roll off the X-ray flux at
$z < 0.05$.  To the extent that faint (non-AGN)  X-ray sources contribute 
significantly to the X-ray background this results in an underestimate 
of $\eta_c$.
We take the lower limit of the $\eta_c$ integral to be $z=0.01$ ($30h^{-1}Mpc$).  This
has the effect of a flux cut for the power law models but in any case it has
little effect on the integral.

Figure 8 is a plot of the computed $\eta_c (0)$ as a function of the correlation
scale length $r_\circ$ for parameters of both the GB6-PMN and FIRST data sets.  Although 
the predicted number densities, $N(z)$, of the Dunlop \& Peacock model are 
within $15\%$ of the observed values ($1.50/deg^2$ for GB6-PMN and $57.3/deg^2$ 
for FIRST), the models were renormalized to agree with the observed values.
All other parameters were taken from the ``best guess'' model, i.e. $\gamma=2.0$,
$\epsilon=1$, Einstein-de Sitter universe, Comastri et al. X-ray emissivity
model (roughly equivalent to $q=3.5$).  The horizontal lines in Figure 8 are
the amplitudes of the observed CCFs got from fitting the $W(\theta)$s of \S3 to
the functional form of $K_\gamma (\theta)$. The formal errors in these fits are on the
order of $20\%$; however, recall that in the case of GB6-PMN the $\chi^{2}$ of the
fit was not good.  The implications are discussed in \S5.

The uncertainties in the model curves were not indicated in Figure 8 because
they are not well known.  However we now discuss how varying parameters
quantitatively changes the curves.  In all cases, except for varying 
$\gamma$, the curves are simply displaced vertically.   If one substitutes 
the power law evolution model for the Comastri et al. model for the XRB 
the values for $\eta_c$ are changed by factors of $2.5$, $1.4$, and $0.7$ 
for $q=2$, $3$, and $4$ respectively.  Although we have argued that linear
clustering growth is appropriate for the current analysis, for stable 
clustering evolution, i.e. $\epsilon=0$, $\eta_c$ is larger by a 
factor of $\sim 1.5$.  In the extreme case,
non-evolving clustering in the comoving frame, i.e., $\epsilon=-1$,
$\eta_c$ is increased by a factor of $\sim 2.6$.

A change in the value of $\gamma$ in our ``best guess'' model changes the
slope as well as the amplitude  of the model curves in Figure 8.  If 
$\gamma=1.8$, the value observed for nearby galaxies (\cite{peb80}), the
value of $\eta_c (0)$ for $r_\circ = 5h^{-1}Mpc$ is a factor of $\sim 1.3$ 
larger than in Figure 8.  Because of the different scale length dependence,
there is an $r_\circ$ above which the modified $\eta_c (0)$ will be less than
the ``best guess'' value.  This value is $16h^{-1}Mpc$.  
If in addition one includes stable clustering the multiplicative factor is 
$\sim 1.8$ and the corresponding crossover $r_\circ$ is $100h^{-1}Mpc$.

To access the dependence of $\eta_c$ on the radio luminosity function
we have recomputed $\eta_c$ for the density/luminosity evolution model of
Dunlop \& Peacock (1990).  This and the previous free-form model represent the
spread in the models considered by Dunlop \& Peacock. The dependence is not
large.  For the GB6-PMN parameters $\eta_c$ decreases by a factor  $0.92$ while
for the FIRST parameters $\eta_c$ increases by a factor of $1.1$.  Both of these
radio luminosity functions predict a large number of low luminosity 
($\leq 10^{30}erg s^{-1} Hz^{-1}$) radio 
sources and the value of $\eta_c$ contains a non-neglible contribution from
these sources.  The contribution to $\eta_c$ from radio sources with
luminosities $\nu L_\nu \le 3 \times 10^{39}erg s^{-1}$  is $6\%$
for the GB6-PMN data and $30\%$ for the FIRST data.  The possible
significance of this rather large contribution will be discussed in \S5.

Finally, we investigated the dependence of the analyses on the large-scale
geometry of the universe.  For open ($\Lambda=0$) universes, the value of
$\eta_c$ increases somewhat, a factor of $\sim 1.3$ for an $\Omega_\circ =0.1$
universe and a factor of $\sim 1.1$ for an $\Omega_\circ =0.3$ universe.
For flat, lambda universes, the factors are $\sim 0.83$ for the 
$\Omega_\circ =0.1$ universe and $\sim 0.85$ for the $\Omega_\circ =0.3$ universe.

\section{Discussion}
We begin by considering the fits of our ``best guess'' model to the two data
sets as indicated in Figures 1 and 2.  For the FIRST data this implies that 
$W (0)=\eta_c(0)/\langle N\rangle \langle I\rangle = 4.9 \pm 0.9 \times 10^{-4}$
or $\eta_c(0) = 2.4 \pm 0.4 \times 10^{-9} erg s^{-1} cm^{-2} sr^{-1}$ where
$W (0)$ is the fitted amplitude of the angular CCF.
This value is indicated by a horizontal line in Figure 8 and implies a
cross-correlation scale length $r_\circ$ of  $5.7 \pm 0.5 h^{-1}$ $Mpc$ where
the error is the statistical error of the fit.  It was found in \S4 that varying
the model parameters from the ``best guess'' values most often results in an
increase in the predicted value of $\eta_c (0)$ and, therefore, a decrease in the value 
of $r_\circ$ inferred from the observed CCF.  For power
law models of the evolution of the X-ray emissivity, only for $q \ge 3.5$ does
the predicted value of $r_\circ$ exceed the value implied by the ``best guess''
model and such models imply a local X-ray emissivity below that observed.
Therefore, we consider that, with two caveats, $5.7 \pm 0.5 h^{-1}$ $Mpc$ is an upper 
limit to the cross-correlation scale length.  The first caveat is that the
Dunlop-Peacock radio luminosity function does not seriously overestimate 
the number of low luminosity sources.  If it does then the predicted
$\eta_c$ will decrease and the implied scale length increase accordingly.
The other caveat is that the universe has a vanishing
cosmological constant.  In a flat, low $\Omega_\circ $ universe the implied
$r_\circ$ is increased by about $10\%$.  

If the Poisson term, $\eta_P$, makes a significant contribution to the FIRST/X-ray
CCF then the implied value of $\eta_c$ is smaller which in turn lowers 
the estimate of $r_\circ$.  Suppose that half the amplitude of the observed
$W(0)$ is due to Poisson fluctuations.  Correcting for these fluctuations and
fitting the clustering term to the residuals implies a clustering amplitude of
$\eta_c(0) = 1.9 \pm 0.4 \times 10^{-9} erg s^{-1} cm^{-2} sr^{-1}$ and 
a correlation scale length of $5.0 \pm 0.5 h^{-1}$ $Mpc$.  It seems unlikely
that the Poisson contribution could be more than this and still be consistent 
with the observed $W(\theta)$.

The fitted amplitude of $\eta_c (\theta)$ to the GB6-PMN data implies
$r_\circ = 10 \pm 2 h^{-1}$ $Mpc$; however, as pointed out in \S4 the $W(\theta)$
profile of this data indicates that Poisson fluctuations dominate and, 
therefore, this value is clearly an overestimate.  Assuming
that the observed CCF is entirely due to Possion fluctuations, the fitted
amplitude of the angular CCF is (see Figure 1) $W (0) = 1.7 \pm 0.3 
\times 10^{-3}$.  From Equation (4-11)
\begin{equation}
W (0) = {\eta_P (0)\over \langle N\rangle \langle I\rangle} = 
{B(0) \bar{I_r} \over \langle N\rangle \langle I\rangle}
\end{equation}
where $\langle I\rangle$ is the mean intensity of the X-ray background and $\bar{I_r}$
is the mean X-ray intensity of the radio sources in the survey.  Therefore, the fraction
of the XRB that is accounted for by the survey radio sources is given by
\begin{equation}
{\bar{I_r} \over \langle I\rangle} = {W (0) \langle N\rangle \over B(0)}.
\end{equation}
Substituting the inferred value of $W (0)$ in this expression implies that
$2.7\%$ of the XRB is due to GB6-PMN radio sources.  This is, of course, an overestimate
because clustering has not been taken into account.  A better
estimate is got by assuming the GB6-PMN/X-ray data
has the same normalized clustering as that of the FIRST/X-ray data, i.e.
$W(0) = 4.9 \times 10^{-4}$.  Subtracting this from the GB6-PMN/X-ray CCF and fitting
the residuals to $\eta_P$ gives a Poisson amplitude of $W_P (0) = 1.3 \times 10^{-3}$
which implies $2.1\%$ of the XRB is due to GB6-PMN radio sources.  It seems unlikely that
the fraction of radio source contribution to the X-ray background could be much less
than $2\%$ without requiring a much larger clustering contribution than 
is allowed by the observed $W(\theta)$ profile.  We note in passing that under the assumptions 
that the radio number counts have a Euclidean dependence on radio flux S, i.e.,
$N(>S) \propto S^{-{3 \over 2}}$, and that the average X-ray to radio luminosity
is independent of redshift, the contribution of radio sources to the X-ray background
saturates at $S = 20 \mu Jy$.  We hasten to add, however, that neither of these assumptions
is likely to be true.  

The profile of the FIRST/X-ray CCF is consistent with no 
Poisson fluctuations.  However, if we assume that half the amplitude
$W (0)$ of the observed CCF is due to Poisson fluctuations then the implied contribution
of the FIRST sources to the XRB is $\sim 20\%$.  We consider this to be an upper limit.

Treyer and Lahav (1996) suggested that cross-correlation analyses of the type above might
enable one to map the X-ray volume emssivity as a function of redshift.  Unfortunately,
the current result is not very useful in this regard.  Although the mean redshift of
the two radio surveys is quite large, $z \sim 1$, the primary contributions to the
CCF comes from lower redshifts, i.e. $\sim 50\%$ from $z < 0.2$.  The result is that
the CCF analysis is not very senstive to the evolution of emissivity.  For example,
the $\eta_c$ for a model with non-evolving emissivity, 
$\varepsilon_b = 9.6 \times 10^{38}h$ $erg~s^{-1} Mpc^{-3}$, is the same as
the $\eta_c$ for our ``best guess'' model.  Although such a model is wildly 
inconsistent with the level of the XRB, it is quite consistent with both the observed
CCF and local X-ray emissivity (\cite{miy94}).

Thus far only upper limits on the clustering of the hard X-ray background have 
appeared in the literature.  We are left with the question of how to interpret the
cross-correlation reported in this paper.  If one assumes  ``linear biasing''
then
\begin{equation}
{\delta \rho \over \rho} = {\delta n_r \over b_r n_r} = 
{\delta \varepsilon_x \over b_x \varepsilon_x}
\end{equation}
where $\rho$ is mass density, $n_r$ is radio source density, $\varepsilon_x$
is X-ray volume emissivity, $\delta$ indicates $rms$ fluctuations in these
quantities, and $b_r$ and $b_x$ are the radio and X-ray bias factors.  To
the extent that the bias factors are independent of scale, the spatial ACFs
of the quantities are related by
\begin{equation}
\xi _m (r) = \xi _r (r)/{b_r}^2 = \xi _x (r)/{b_x}^2 .
\end{equation}
If $\xi (r) \propto ({r \over r_\circ})^{-\gamma}$, then 
$r_{\circ ,r} \propto b_r ^{2\over \gamma}$ and $r_{\circ ,x} \propto 
b_x ^{2\over \gamma}$.  Then the
cross-correlation function satisfies
\begin{equation}
\xi _{rx} = {\langle \delta n_r \delta \varepsilon_x \rangle \over 
n_r \varepsilon_x} \propto b_r b_x \propto (r_{\circ ,r}r_{\circ ,x})
^{\gamma \over 2}.
\end{equation}
Assuming that $\xi_{rx} \propto ({r \over r_{\circ ,xr}})^{-\gamma}$, the
cross-correlation scale length, $r_{\circ ,xr}$ is equal to the 
geometric mean of the two auto-correlation scale lengths, i.e.,
\begin{equation}
r_ {\circ ,xr} = (r_{\circ ,r}r_{\circ ,x})^{1 \over 2}.
\end{equation}
While it is likely that both of these assumptions are violated to some extent,
it seems reasonable that equation (5-6) is a valid approximation.

The expression for the unnormalized ACFs for X-ray flux and radio source counts
is equation (4-15) with ${\mathcal{F}}(z) N(z)$ replaced by either
${\mathcal{F}}(z)^2$ for the X-ray ACF or $N(z)^2$ for the radio ACF.  If the
PSF is a delta function and the cell size is small, the expression for 
$K_\gamma (\theta)$ becomes $K_\gamma (\theta) = \theta ^{1-\gamma}\omega ^2$ 
(\cite{tl96}) where $\omega $ is the solid angle of a cell.
Then
\begin{equation}
\eta_{ACF} = H_\gamma r_\circ ^{\gamma} \theta ^{1-\gamma} \omega ^2 \int f(z)^2
(1+z)^{\gamma -{3 \over 2} -\epsilon} {[1-(1+z)^{-{1 \over 2}}]}^{1-\gamma} dz
\end{equation}
where $f(z)$ is either $N(z)$ or ${\mathcal{F}}(z)$.  To obtain the normalized 
ACFs one must divide by either $\langle N \rangle ^2$ or $\langle I \rangle ^2$
where $\langle N \rangle$ or $\langle I \rangle =
\omega \int f(z)dz$.  Thus
\begin{equation}
W_{ACF} (\theta) = H_\gamma r_\circ ^\gamma \theta ^{1-\gamma}
{\int f(z)^2(1+z)^{\gamma -{3 \over 2} -\epsilon}
[1-(1+z)^{-{1 \over 2}}]^{1-\gamma}dz \over (\int f(z)dz)^2}.
\end{equation}

The strongest limit on the $2-10~keV$ ACF is from Carrera et al. (1993) where
the $2\sigma$ limit at $\theta = 2^\circ$ is $W_x (2^\circ ) < 5 \times 10^{-4}$.
Substituting this value into equation (5-8) and using our ``best guess'' model
for the XRB we find that $r_{\circ ,x} < 7h^{-1}~Mpc$.  Cress et al. (1996)
have recently measured the radio ACF for the intial FIRST data release (about
half the number of sources used in this paper).  From their Figure 1 we find that 
$W_r (2^\circ ) \sim 2 \times 10^{-3}$.  Again substituting this value and the
``best guess'' model parameters into equation (5-8) we find that 
$r_{\circ ,r} \sim 10h^{-1}~Mpc$.  This value agrees with their preliminary reported
value (\cite{cre96}).
Substituting these values of $r_{\circ ,x}$
and $r_{\circ ,r}$ into equation (5-6) implies a constraint on the
cross-correlation length of $r_{\circ ,xr} \simle 8h^{-1}~Mpc$ which is consistent 
with our observed value of $\simle 6h^{-1}~Mpc$.  On the other hand, substituting 
$r_{\circ ,r} \approx 10h^{-1}~Mpc$ and $r_{\circ ,xr} \simle 6h^{-1}~Mpc$
into equation (5-6) yields $r_{\circ ,x} \simle 4h^{-1}~Mpc$ which is
smaller than the correlation length of galaxies.  This is somewhat
bothersome and may imply that either we have underestimated the
cross-correlation length scale or that $r_{\circ ,r} \approx 10h^{-1}~Mpc$
is an overestimate.  The latter will undoubtedly be clarified as more of
the FIRST survey is completed.

Finally, we pointed out in \S4 that the estimate of $r_{\circ ,xr}$ 
would be increased if low luminosity radio sources are significantly
overestimated by the Dunlop-Peacock models.  However, the implied
constraint on $r_{\circ ,x}$ is relatively insensitive to the
luminosity function.  For example, if we artificially cut off the
radio luminosity function at $\nu L_\nu = 10^{40} erg~s^{-1}$
the inferred limit of $r_{\circ ,xr}$ increases to
$\simle 8h^{-1}~Mpc$ while $r_{\circ ,r}$ becomes $15h^{-1}~Mpc$.  Then
equation (5-6) still implies that $r_{\circ ,x} \simle 4h^{-1}~Mpc$.

\section{Conclusions}
The $2-10~keV$ X-ray background (at $3^\circ$ angular resolution) is
signficantly correlated with both $1.4GHz$ FIRST radio source counts
and $5GHz$ GB6 and Parkes-MIT-NRAO radio source counts.  The amplitude
of the cross-correlation functions for these two data sets is
$W_{xr}(0) = 4.9 \pm 0.9 \times 10^{-4}$ for the FIRST/X-ray CCF and
$W_{xr}(0) = 1.7 \pm 0.3 \times 10^{-3}$ for the GB6-PMN/X-ray CCF.
Interpreted in terms of a ``best guess'' model ($\Omega_\circ = 1$,
linear growth of perturbations, Dunlop-Peacock radio luminosity function,
and a unified AGN model of the XRB), the FIRST/X-ray CCF implies a
comoving correlation length of $r_{\circ ,xr} = 5.7 \pm 0.5~h^{-1}~Mpc$ 
(statistical error only).  The dependence of this value on model parameters
indicates that a reasonable upper limit to the correlation length is
$r_{\circ ,xr} \simle 6h^{-1}~Mpc$.  If the FIRST ACF correlation length
is $r_{\circ ,r} \approx 10h^{-1}~Mpc$ as has been reported, then the
implied XRB ACF correlation length is $r_{\circ ,x} \simle 4h^{-1}~Mpc$
which is somewhat smaller than the galaxy-galaxy correlation 
length, $5h^{-1}~Mpc$.  We note in passing that if 
$r_{\circ ,r} \sim 7h^{-1}~Mpc$ then the implied X-ray correlation length
is $r_{\circ ,x} \sim 5h^{-1}~Mpc$ and we suggest that future radio 
observations may reveal the smaller ACF implied by this value.  In
any case, a low value of the X-ray correlation length,
$\simle 5h^{-1}~Mpc$, is consistent with the hypothesis that a significant
fraction of the XRB is due to objects which are less strongly clustered than
luminous AGNs, i.e. QSOs.

The GB6-PMN/X-ray CCF is dominated by Poisson noise and can be
used to infer that $\sim 2\%$ of the $2-10~keV$ background is due to
$5GHz$ radio sources with fluxes in excess of $50mJy$.

\acknowledgements
I would like to acknowledge useful conversations with Rob Crittenden,
Ofer Lahav, Neil Turok, Insu Yi, and especially with Daniel Eisenstein.
Keith Jahoda kindly supplied me with 
the HEAO I A2 X-ray data and several data handling programs.  Much of 
this work was completed at Princeton University where I benefitted 
greatly from the subroutines of Ed Groth.  This work was supported in
part by NASA grant NAG 5-3015, NSF grant PHY-9222952, and the
Monel Foundation.

\begin{figure}[phbt]
\centerline{\rotate[r]{\vbox{\epsfysize=17cm\epsfbox{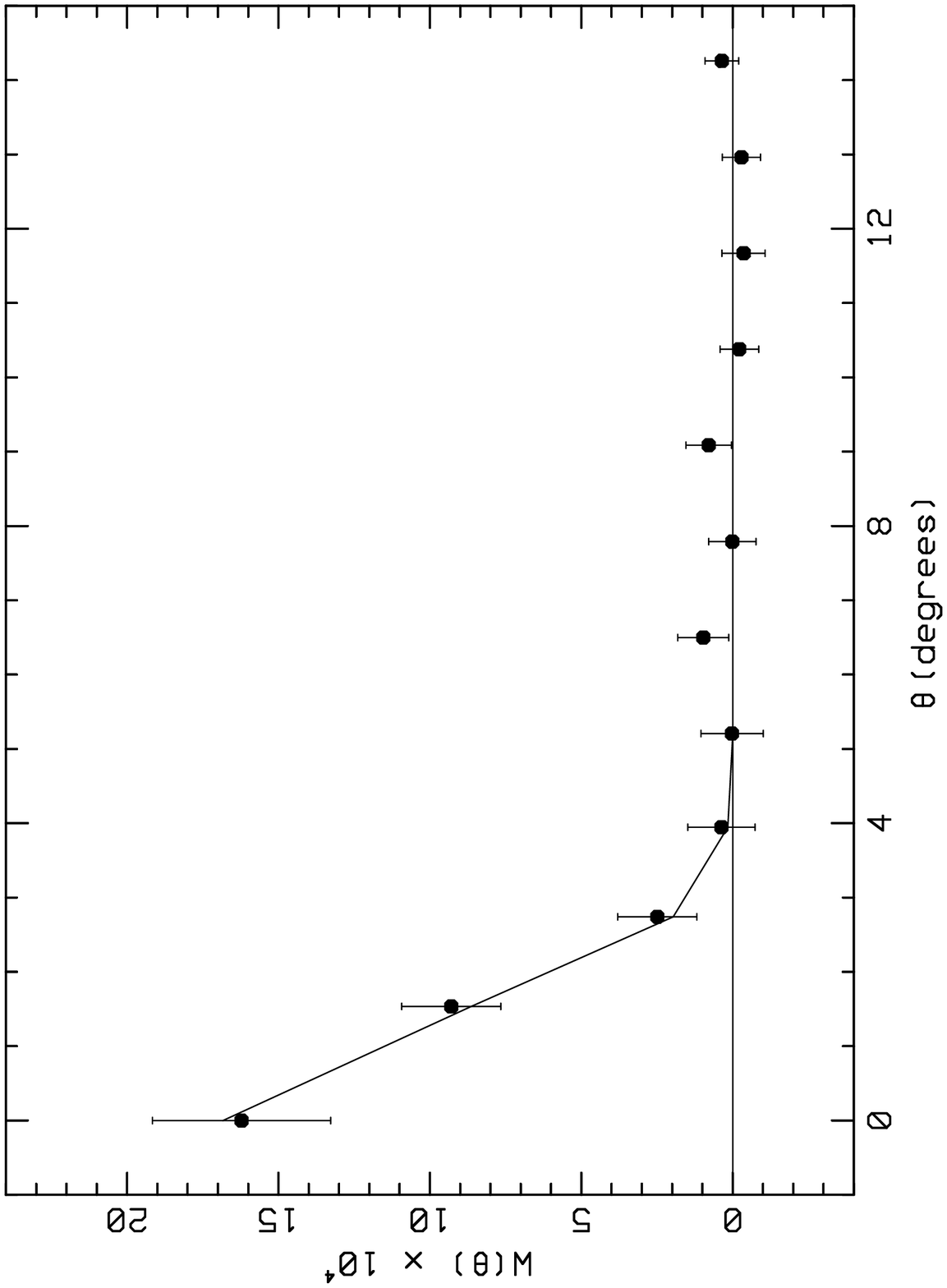}}}}
\caption[]
{Cross-correlation function of diffuse $2-10~keV$ X-rays with the $5 GHz$ GB6-PMN radio
surveys.  The errors are statistical only and are highly correlated.  The curve is a 
fit to the profile expected for Poisson fluctuations convolved with the X-ray beam.  
See \S4 .
}
\end{figure}

\begin{figure}[phbt]
\centerline{\rotate[r]{\vbox{\epsfysize=17cm\epsfbox{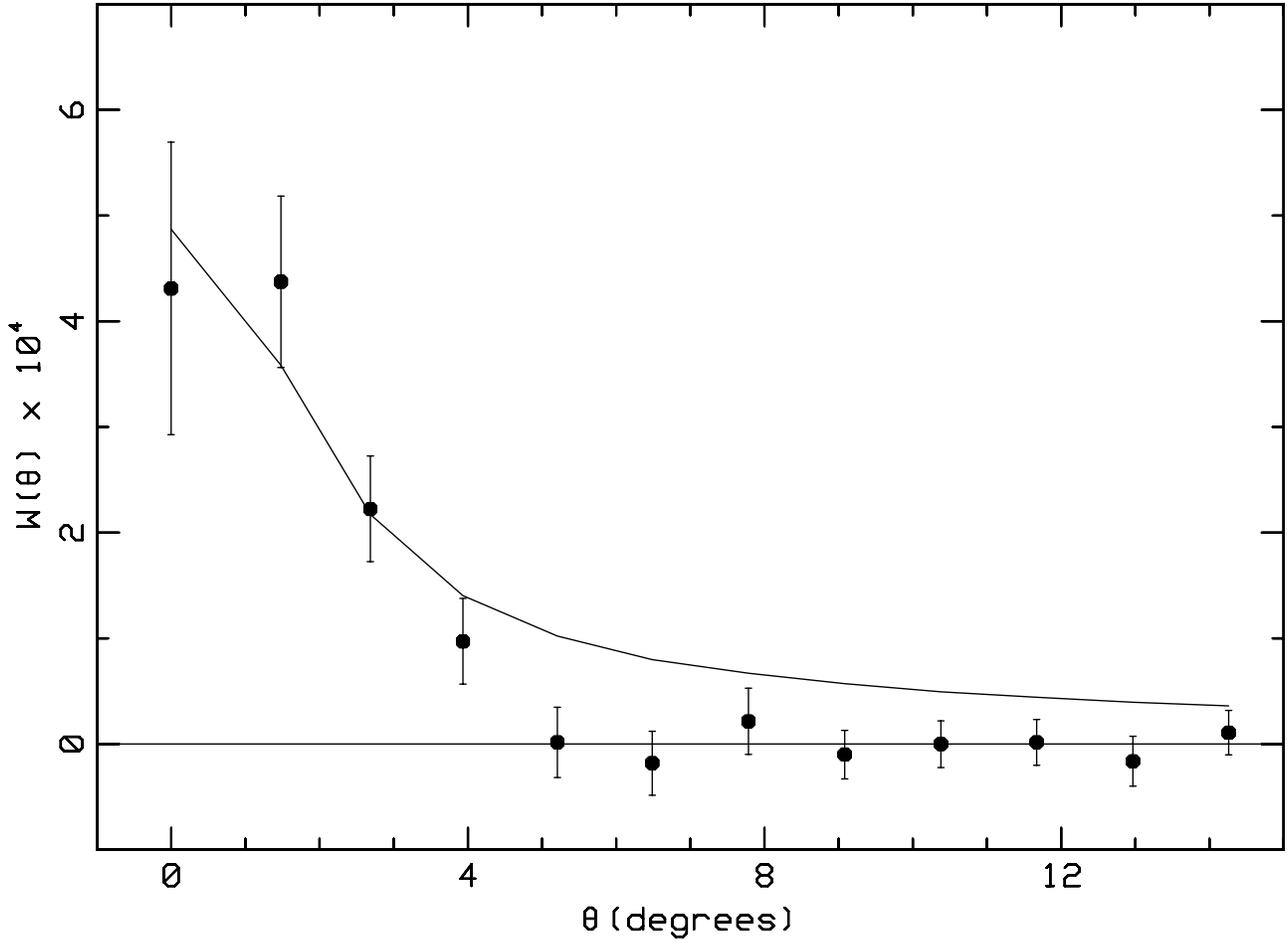}}}}
\caption[]
{Cross-correlation function of diffuse $2-10~keV$ X-rays with the $1.4 GHz$ FIRST radio
survey.  The errors are statistical only and are highly correlated.  The curve is a 
fit to the profile expected for $\gamma = 2$ spatial clustering.  See \S4 .
}
\end{figure}

\begin{figure}[phbt]
\centerline{\rotate[r]{\vbox{\epsfysize=17cm\epsfbox{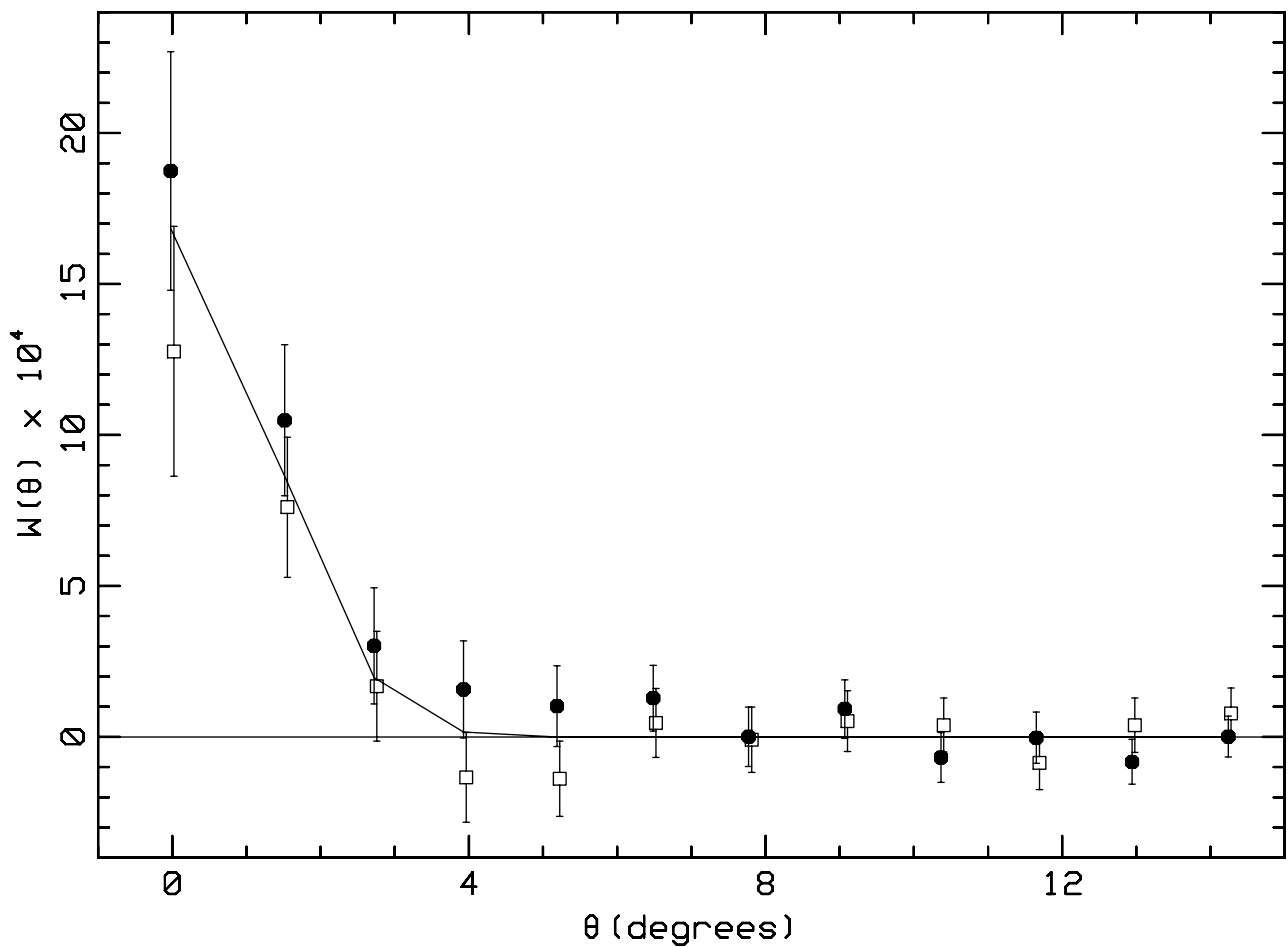}}}}
\caption[]
{Cross-correlation function of diffuse $2-10~keV$ X-rays with the $5 GHz$ GB6 radio
survey (filled circles) and PMN survey (open squares).  The errors are statistical 
only and are highly correlated.  The curve is the same fit as in Figure 1.
}
\end{figure}

\begin{figure}[phbt]
\centerline{\rotate[r]{\vbox{\epsfysize=17cm\epsfbox{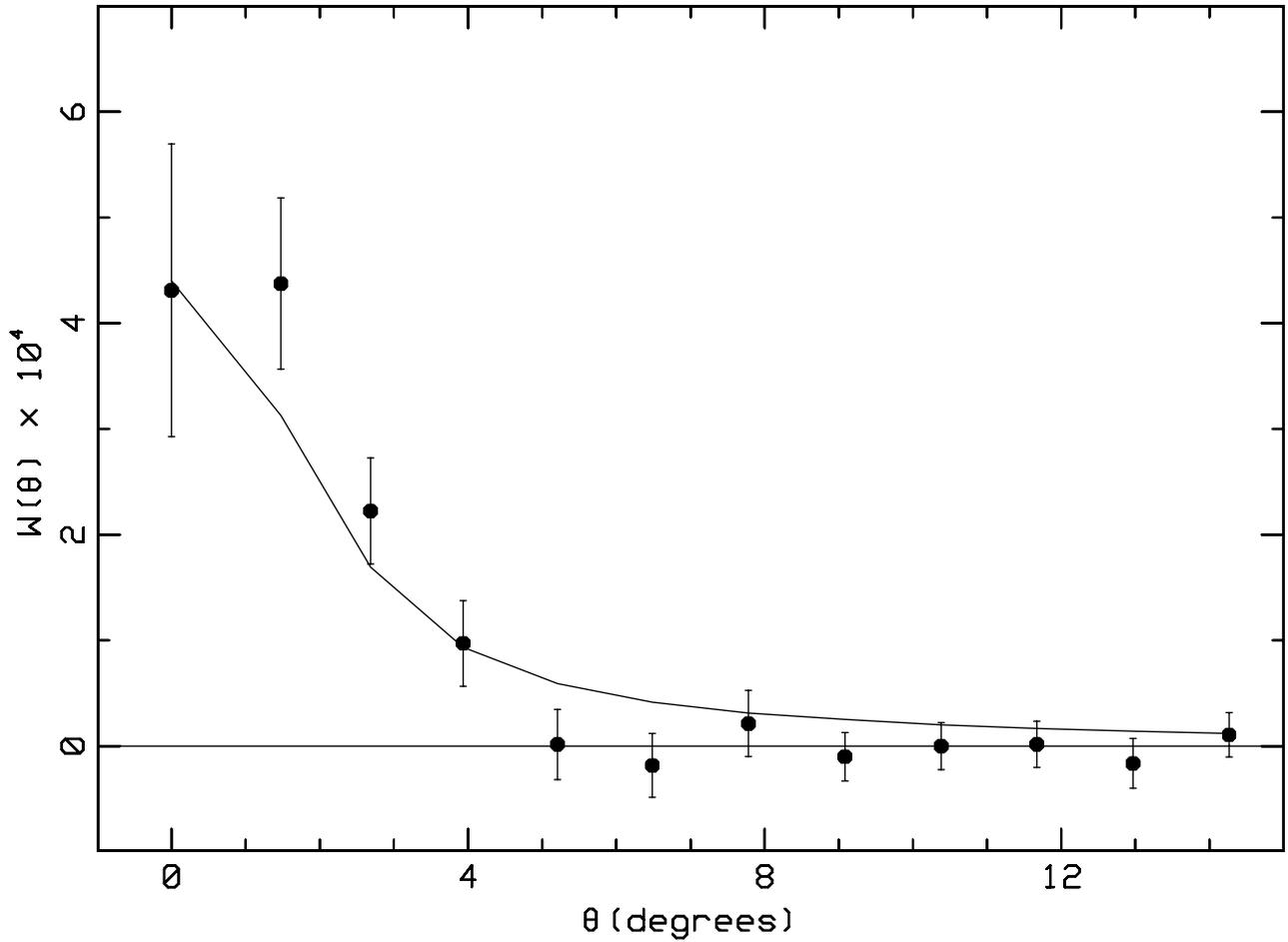}}}}
\caption[]
{Cross-correlation function of diffuse $2-10~keV$ X-rays with the $1.4 GHz$ FIRST radio
survey.  The data is the same as in Figure 2.  The curve is the fit to the profile
expected for $\gamma = 2$ spatial clustering cut off at a physical distance of
$30h^{-1}Mpc$.  See \S4.
}
\end{figure}

\begin{figure}[phbt]
\centerline{\rotate[r]{\vbox{\epsfysize=17cm\epsfbox{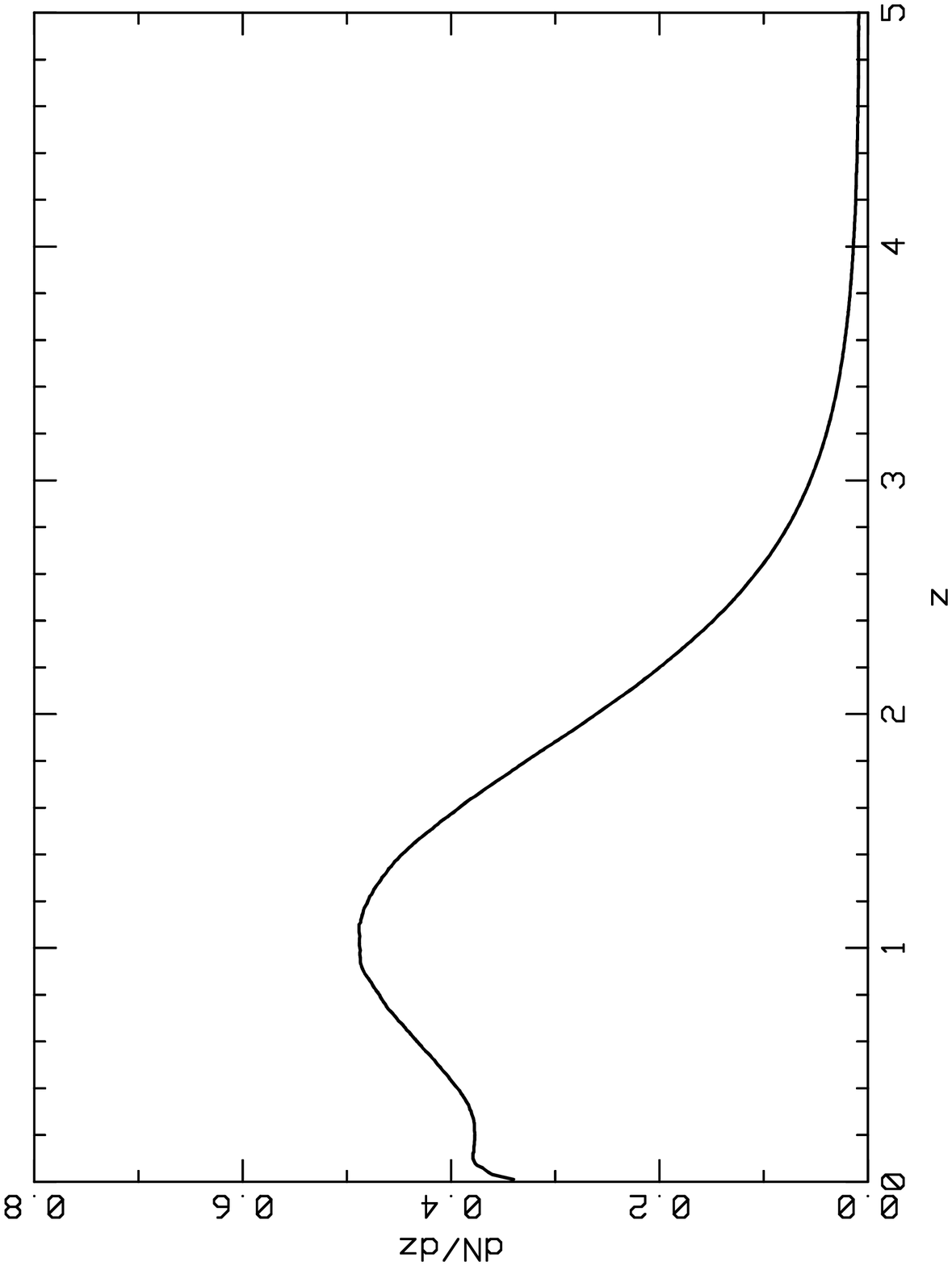}}}}
\caption[]
{The redshift distribution of $50 mJy$ flux limited $5 GHz$ radio number counts
predicted by the luminosity function model (model 1 MEAN-z) of Dunlop \& Peacock
(1994).
}
\end{figure}

\begin{figure}[phbt]
\centerline{\rotate[r]{\vbox{\epsfysize=17cm\epsfbox{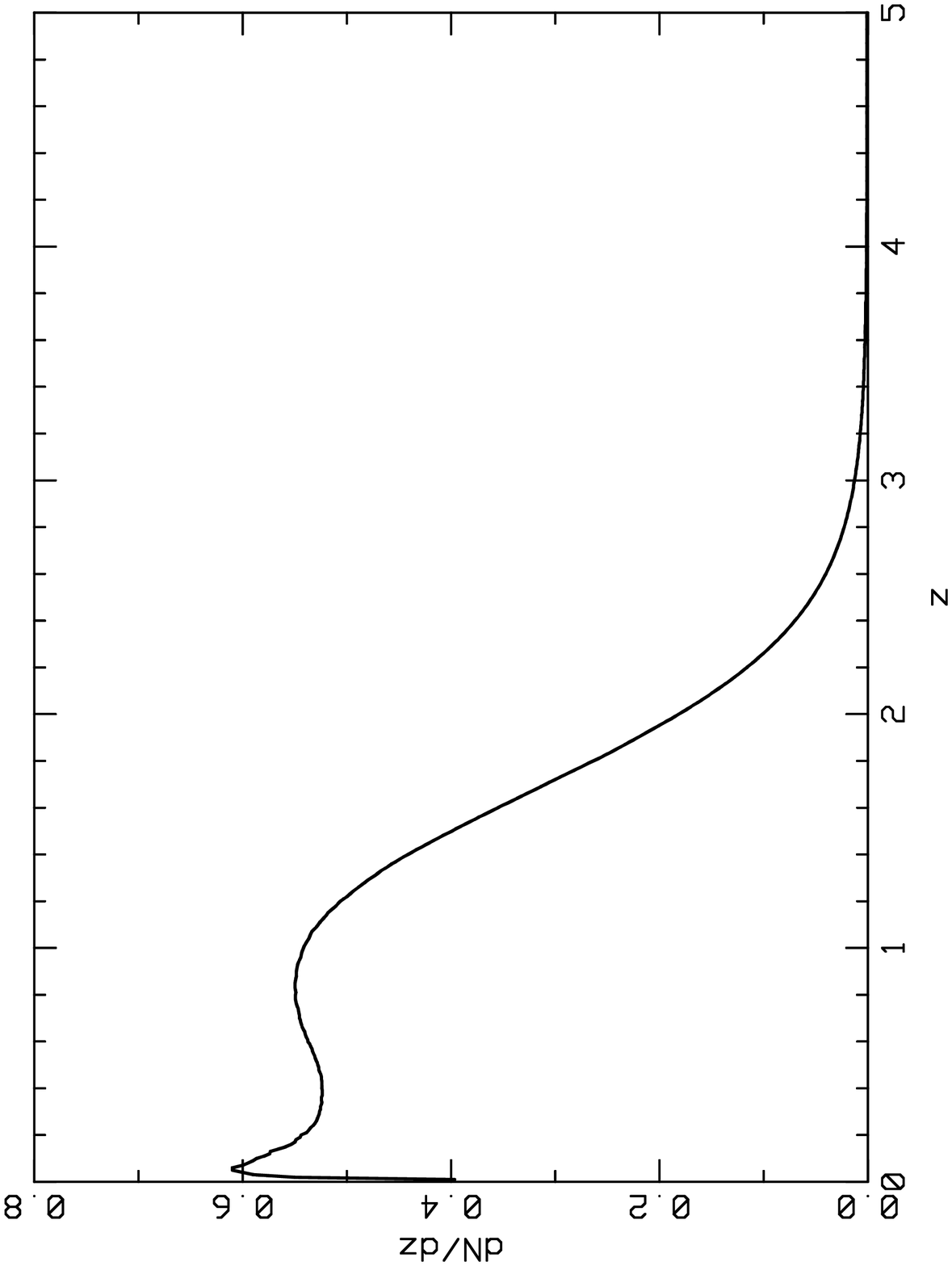}}}}
\caption[]
{The redshift distribution of $1.5 mJy$ flux limited $1.4 GHz$ radio number counts
predicted by the luminosity function model (model 1 MEAN-z) of Dunlop \& Peacock
(1994).
}
\end{figure}

\begin{figure}[phbt]
\centerline{\rotate[r]{\vbox{\epsfysize=17cm\epsfbox{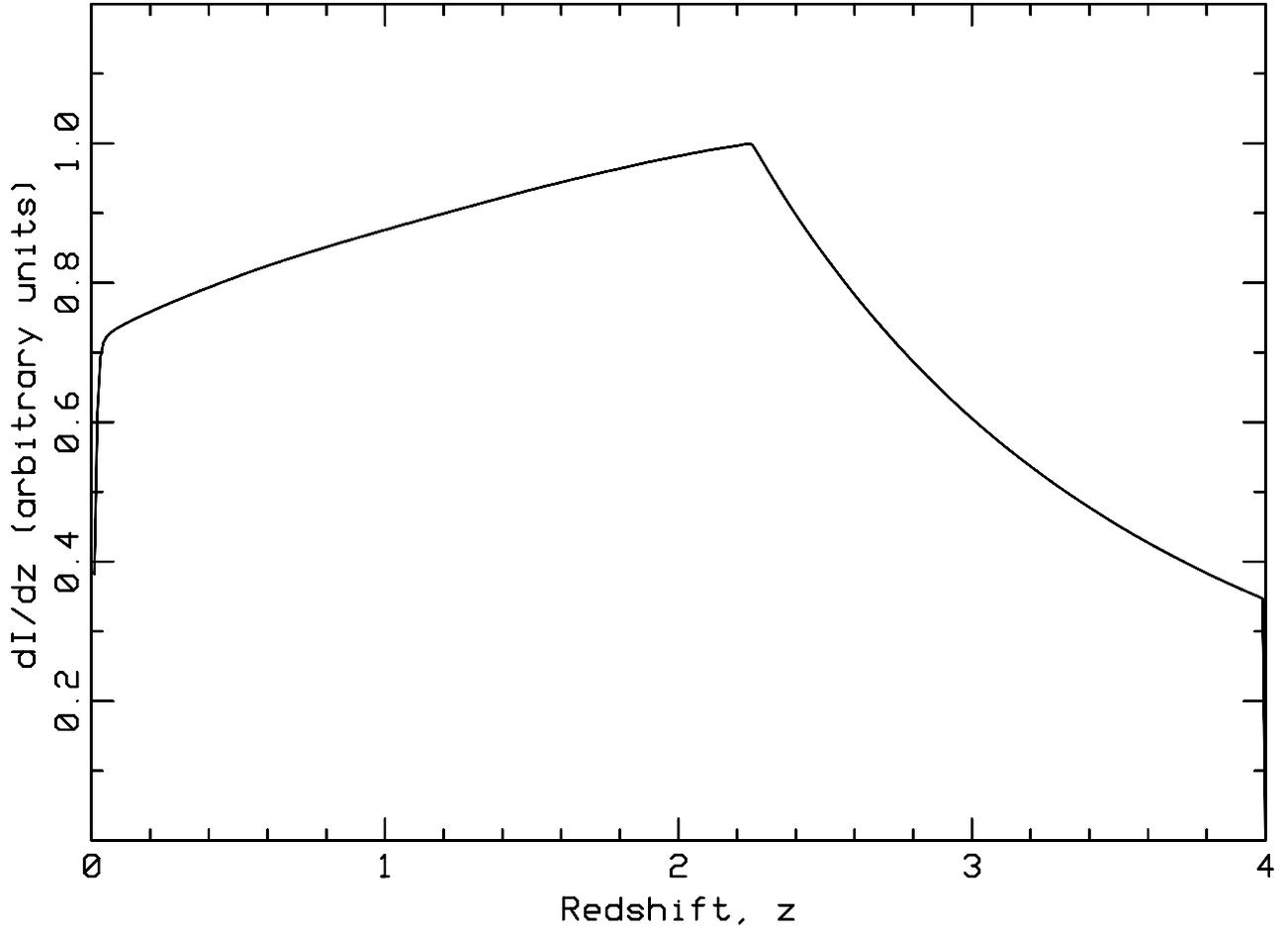}}}}
\caption[]
{Redshift distribution of the $2-10~keV$ X-ray Background, $dI/dz$,
from the unified AGN model of Comastri et al. (1995).  Sources with
fluxes exceeding $3\times 10^{-11}~erg~s^{-1}~cm^{-1}$ have been cut.
}
\end{figure}

\begin{figure}[phbt]
\centerline{\rotate[r]{\vbox{\epsfysize=17cm\epsfbox{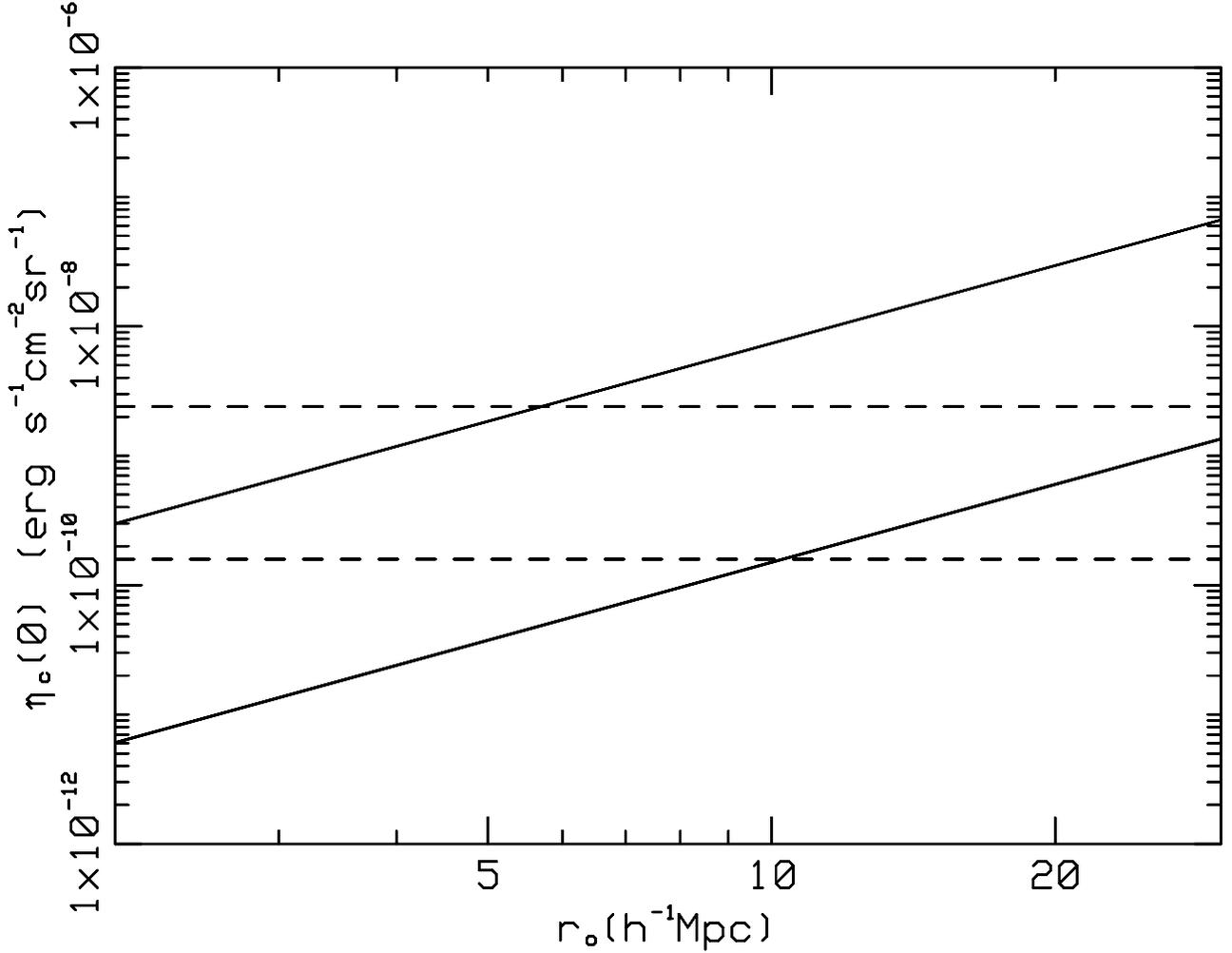}}}}
\caption[]
{Predicted cross-correlation amplitude, $\eta_c (0)$, as a function of correlation
scale length, $r_\circ$, using the ``best guess'' model discussed in the text.
The upper solid curve is that predicted for the FIRST/X-ray CCF and the lower solid curve for
the GB6-PMN/Xray CCF.  The upper dashed line corresponds to the amplitude fitted 
to the FIRST/X-ray data and the lower dashed line to the amplitude fitted
to the GB6-PMN/X-ray data.  Note: the dashed lines should be considered upper limits
since Poisson fluctuations have not been corrected for.  See \S4 and \S5.
}
\end{figure}


\begin{thebibliography}{}
\bibitem[Allen, Jahoda \& Whitlock 1994]{ajw94}Allen, J., Jahoda, K., \&
	Whitlock, L. 1994, Legacy, 5, 27
\bibitem[Almaini \& Fabian 1997]{af97}Almaini, O., and Fabian, A. C. 1997, MNRAS,
	in press (astro-ph/9704128)
\bibitem[Almaini et al. 1997]{alm97}Almaini, O., Shanks, T., Griffiths, R. E.,
	Boyle, B. J., Roche, N., Georgantopoulos, I., \&  Stewart, G. C 1997, 
	MNRAS, in press (astro-ph/9704117)
\bibitem[Barcons, Fabian, \& Carrera 1997]{bfc97}Barcons, X., Fabian, A. C., \&
	Carrera, F. J. 1997, MNRAS, in press
\bibitem[Becker, White \& Helfand 1995]{bwh95}Becker, R. H., White, R. L., \& 
	Helfand, D. J. 1995, ApJ, 450, 559
\bibitem[Bennett et al. 1996]{ben96}Bennett, C. L. et al. 1996, ApJ, 464, L1
\bibitem[Boldt 1987]{bol87}Boldt, E. 1987, Phys Rept, 146, 215
\bibitem[Boughn, Crittenden, \& Turok 1997]{bct97}Boughn, S. P., Crittenden, R. G.,
	Turok, N. G. 1997, submitted to ApJ
\bibitem[Boyle et al. 1994]{boy94}Boyle, B. J., Giffiths, R. E., Shanks, T., 
	Stewart, G. C., \& Georgantopoulos, I. 1994, MNRAS, 271, 639
\bibitem[Comastri et al. 1995]{com95}Comastri, A., Setti, G., Zamorani, G., \&
	Hasinger, G. 1995, A \& A, 296, 1
\bibitem[Cress et al. 1996]{cre96}Cress, C. M., Helfand, D. J., Becker, R. H.,
	Gregg, M. D., \& White, R. L. 1996, ApJ, 473, 7
\bibitem[Dunlop \& Peacock]{dp90}Dunlop, J. S. \& Peacock, J. A. 1990, MNRAS, 247, 19
\bibitem[Gendreau et al. 1995]{gen95}Gendreau, K. C., et al. 1995, PASJ, 47, L5
\bibitem[Georgantopoulos et al. 1997]{geo97}Georgantopoulos, I., Stewart, G. C.,
	Blair, A., J., Shanks, T., Griffiths, R. E., Boyle, B. J., Almaini, O., \&
	Roche, N. 1997, MNRAS, in press
\bibitem[Gregory et al. 1996]{gsdc96}Gregory, P. C., Scott, W. K., Douglas, K.,
	\& Condon, J. J. 1996, ApJS, 103, 427
\bibitem[Griffith et al. 1994]{gri94}Griffith, M. R., Wright, A. E., Burke, B. F.,
	Ekers, R. D. 1994, ApJS, 90, 179
\bibitem[Griffith et al. 1995]{gri95}Griffith, M. R., Wright, A. E., Burke, B. F.,
	Ekers, R. D. 1995, ApJS, 97, 347
\bibitem[Fisher et al 1994]{fis94}Fisher, K. B., Davis, M., Strauss, M. A., 
	Yahil, A., \& Huchra, J. 1994, MNRAS, 266, 50
\bibitem[Hasinger et al. 1993]{has93}Hasinger, G., Burg, R., Giacconi, R.,
	Hartner, G., Schmidt, M., Trumper, J., \& Zamorani, G. 1993, A \& A, 275, 1
\bibitem[Jahoda 1993]{jah93}Jahoda, K. 1993, Adv Space Res, 13, No 12, 231
\bibitem[Jahoda \& Mushotzky 1989]{jm89}Jahoda, K., \& Mushotzky, R. 1989, ApJ, 346, 638
\bibitem[Lahav, Piran \& Treyer 1997]{lpt97}Lahav, O., Piran, T., \& Treyer, M. A.
	1997, MNRAS, 284, 499
\bibitem[Ling, Frenk \& Barrow 1986]{lfb86}Ling, E. N., Frenk, C. S., \& Barrow, J. D.
	1986, MNRAS, 223, 21
\bibitem[Loan, Wall \& Lahav 1997]{lwl997}Loan, A. J., Wall, J. V., \& Lahav, O. 1997,
	MNRAS, 286, 994
\bibitem[Marshall et al. 1980]{mar80}Marshall, F. E., Boldt, E. A., Holt, S. S., Miller. R. B.,
	Mushotzky, R. F., Rose, L. A., Rothschild, R. E., \& Serlemitsos, P. J. 1980, ApJ, 235, 4
\bibitem[Miyaji et al. 1994]{miy94}Miyaji, T., Lahav, O., Jahoda, K., \& Boldt, E. 1994,
	ApJ, 434, 424
\bibitem[Padmanabhan 1993]{pad93}Padmanabhan, T. 1993, Structure Formation in the Universe,
	(Cambridge University Press, Cambridge)
\bibitem[Peebles 1980]{peb80}Peebles, P. J. E. 1980, The Large-Scale Structure of the
	Universe, (Princeton Univ Press, Princeton, NJ) 
\bibitem[Peebles 1993]{peb93}Peebles, P. J. E. 1993, Principles of Physical Cosmology,
	(Princeton Univ Press, Princeton, NJ)
\bibitem[Piccinotti et al. 1982]{pin82}Piccinotti, G., Mushotzky, R., Boldt, E.,
	Marshall, F., Serlemitsos, P., \& Shafer, R. 1982, ApJ, 253, 485
\bibitem[Refregier, Helfand, \& McMahan 1997]{rhm97}Refregier, A., Helfand, D. J.,
	\& McMahon, R. G. 1997, ApJ, 477,58
\bibitem[Roche et al. 1995]{roc95}Roche, N., Shanks, T., Georgantopoulos, I.,
	Stewart, G. C., Boyle, B. J., \& Griffiths, R. E. 1995, MNRAS, 273, L15
\bibitem[Shafer 1983]{sha83}Shafer, R. A. 1983, PhD Thesis, Univ of Maryland
\bibitem[Treyer \& Lahav 1996]{tl96}Treyer, M. A., \& Lahav, O. 1996, MNRAS, 280, 469
\bibitem[White et al. 1997]{wbhg97}White, R. L., Becker, R. H., Helfand, D. J.,
	\& Gregg, M. D. 1997, ApJ, 475, 479
\bibitem[White \& Stemwedel 1992]{ws92}White, R. A. \& Stemwedel, S. W. 1992, in
	Astronomical Data Analysis Software and Systems I, eds. D. M. Worrall,
	C. Biemesderfer, \& J. Barnes (San Francisco: ASP), 379
\bibitem[Wright et al. 1994]{wri94}Wright, A. E., Griffith, M. R., Burke, B. F.,
	\& Ekers, R. D. 1994, ApJS, 91, 111
\bibitem[Wright et al. 1996]{wri96}Wright, A. E., Griffith, M. R., Hunt, A. J.,
	Troup, E., Burke, B. F., \& Ekers, R. D. 1994, ApJS, 103, 145
\end{thebibliography}
\end{document}